\documentclass[]{article}

\usepackage{graphicx}
\usepackage{fullpage}
\usepackage[normalem]{ulem}
\usepackage{stfloats}
\usepackage{psfrag}
\usepackage[export]{adjustbox}
\usepackage[style=numeric,doi=false,isbn=false,firstinits=true,sorting=none,
                    backend=bibtex]{biblatex}

\newcommand{\un}[1]{\ensuremath{\,\mathrm{#1}}}
\newcommand{\chem}[1]{\ensuremath{\mathrm{#1}}}
\def\imagetop#1{\vtop{\null\hbox{#1}}}

\renewbibmacro{in:}{}

\AtEveryBibitem{\clearfield{month}}
\AtEveryCitekey{\clearfield{month}}
\AtEveryBibitem{\clearfield{day}}
\AtEveryCitekey{\clearfield{day}}
\AtEveryBibitem{\clearlist{abstract}}
\DeclareFieldFormat{url}{{\url{#1}}}
\DeclareFieldFormat{language}{}
\DeclareFieldFormat{file}{}
\DeclareFieldFormat{abstract}{}
\DeclareFieldFormat*{urldate}{}
\begin{document}

\title{X-ray diffraction microscopy based on refractive optics}
\author{Thomas Roth\footnote{now at: European XFEL GmbH, Hamburg, Germany. e-mail: thomas.roth@xfel.eu}, Carsten Detlefs, Irina Snigireva, Anatoly Snigirev\\ \\European Synchrotron Radiation Facility,\\ B.P. 220, 6 rue Jules
Horowitz, 38043 Grenoble Cedex 9, France}
\date{September 21, 2014}
\maketitle

Keywords: Diffraction imaging, X-ray microscopy, Compound refractive lenses, X-ray topography

\begin{abstract}
  We describe a diffraction microscopy technique
  based on refractive optics to study structural variations in
  crystals. The X-ray beam diffracted by a crystal was magnified by
  Beryllium parabolic refractive lenses on a 2D X-ray camera. The
  microscopy setup was integrated into the 6-circle Huber
  diffractometer at the ESRF beamline ID06. Our setup allowed us to
  visualize structural imperfections with a resolution of $\approx
  1\un{\mu m}$. The configuration, however, can easily be adapted for
  sub-$\mathrm{\mu m}$ resolution.
\end{abstract}

\section{Introduction}

X-ray diffraction imaging, known later as X-ray topography, originated
in the late 1920ies and early 1930ies, when researchers revealed the
internal structure of individual Laue spots in diffraction patterns
\cite{Berg31,Barrett31}. To improve the resolution, fine grain
photographic emulsions were exposed and examined under optical
microscopes -- for this reason the technique was sometimes called
X-ray microscopy \cite{Barrett45}. This method was applied for both
mapping of strains in heavily deformed materials such as cold-worked
metals and alloys \cite{Barrett45} and studies of individual
defects in near-perfect single crystals \cite{Ramachandran44}. 

It was assumed that each point on the film or detector corresponds to
a small volume in the reflecting crystal. Simple geometrical optics
then requires the incoming X-ray beam to be tightly collimated, and
the film to be placed as closely as possible to the sample. The
achievable resolution is then limited by the detector resolution, at
best $500\un{nm}$ \cite{Martin06} but more typically $1\un{\mu m}$.

However, in the absence of X-ray optics between the sample and the
film diffraction effects progressively blur the image with increasing
sample-to-detector distance.  For a typical experimental setup
(wavelength $\lambda=1\,\textrm{\AA}$, sample-to-detector distance
$s=1\un{cm}$, and sample feature size $d=1\un{\mu m}$) the diffraction
limited resolution due to propagation of the perturbed wavefront from
the exit surface of the crystal to the detector can be approximated as
\begin{equation}
  \delta
  \approx \frac{\lambda}{d}\cdot s
  = \frac{10^{-10}\un{m}}{10^{-6}\un{m}}\cdot 10^{-2}\un{m}
  = 1\un{\mu m}.
\end{equation}
Note that to image $10\times$ smaller features on the sample
($d=100\un{nm}$ and therefore $\delta=100\un{nm}$) the
sample-to-detector distance would have to be decreased by a factor of
100, $s=100\un{\mu m}$. In most cases this is technically not
feasible. Furthermore, to the best of our knowledge, 2D imaging
detectors with a spatial resolution of $100\un{nm}$ are not yet
available.

On the other hand, conventional X-ray microscopy techniques as
proposed by Kirkpatrick and Baez \cite{kirkpatrick48,baez52} have
been implemented in the hard X-ray domain rather late. Here, an
in-line scheme is used where the beam transmitted through the sample
is magnified by X-ray optics such as mirrors \cite{Underwood86},
Fresnel zone plates \cite{lai95}, Bragg-Fresnel lenses
\cite{Snigirev97}, or refractive lenses \cite{lengeler99}. Such
forward scattering techniques are primarily sensitive to spatial
variations of the X-ray index of refraction which depends mostly on
the local density of the sample.

In this paper we propose a compact scheme for diffraction microscopy
using X-ray refractive lenses between the sample and the detector. The
insertion of refractive optics into the diffracted beam allows
significant improvements of the resolution, potentially 
down to below 100\,nm (a resolution of 300\,nm has been demonstrated 
using a similar lens in transmission X-ray microscopy\cite{Bosak10}). 
Furthermore, the progressive blurring due to the
wavefront propagating from the sample to the detector can be overcome
by a lens, thus reestablishing the direct mapping of intensity
variations on the detector to the reflectivity variations on the
sample. In this case the image resolution can, in principle, reach the
limit imposed by dynamical diffraction effects within the crystal.

Recently, Fresnel zone plates have been used in X-ray reflection
microscopy to image monomolecular steps at a solid surface
\cite{Fenter06} and for scanning X-ray topography of strained silicon
oxide structures \cite{Tanuma06}. CRLs have the advantage that
efficient focusing can be achieved at higher photon energies, $E \gg
10\un{keV}$. Please note that standard KB mirrors are not suited for imaging setups, as they do not fulfill the Abbe-sine condition. More complicated multi-mirror setups are however being developed to overcome this limitation in transmission geometry \cite{matsuyamaHard2012}.

%%%%%%%%%%%%%%%%%%%%%%%%%%%%%%%%%%%%%%%%%%%%%%%%%%%%%%%%%%%%%%%%%%%%%%%

\section{Experimental details}

\begin{figure}[ht!]
  \begin{center}
    \includegraphics[width=0.7\columnwidth, clip=true]{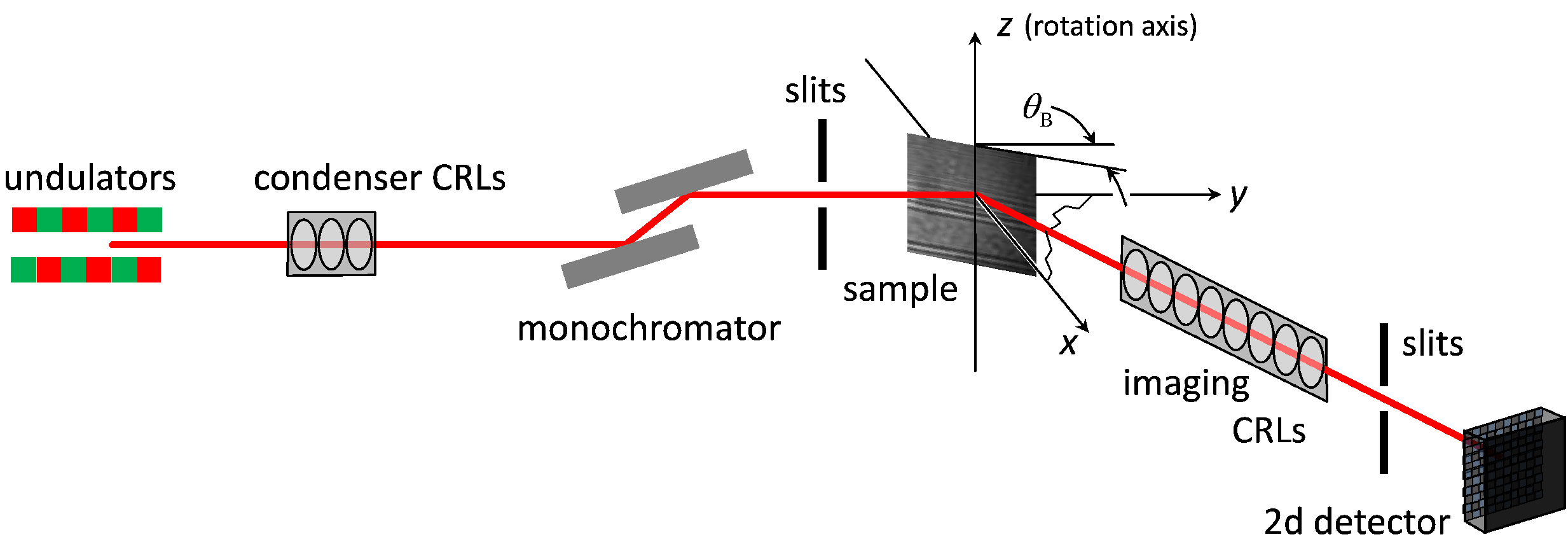}
  \end{center}
  \caption{ \label{fig:setup}Experimental setup for Bragg diffraction
    microscopy. 11\un{keV} X-rays impinge on the sample. The
    diffracted intensity is imaged onto a Sensicam camera via a set of
    66 Beryllium compound refractive lenses (CRLs) with an apex-radius of curvature of
    $50\un{\mu m}$. The scattering plane is horizontal, and the imaged
    features on the sample were aligned parallel to the scattering
    plane.  }
\end{figure}

Our experiment was carried out at the undulator beamline ID06 of the
European Synchrotron Radiation Facility. A cryogenically cooled
permanent magnet in-vacuum undulator \cite{Chavanne09} with a period
of 18\un{mm} and a conventional in-air undulator with a period of
32\un{mm}, combined with a liquid nitrogen cooled Si (111)
monochromator, delivered photons at an energy of 11\un{keV}. A
transfocator located at 38.7\un{m} from the source point (electron beam waist position in the middle between the two undulators) acted as a
condenser, i.e. it focused the photons onto the sample at 67.9\un{m}
distance from the source, using a combination of paraboloid (2D)
compound refractive lenses, CRLs, \cite{lengeler99}: one lens with radius of curvature at the apex $R$= 1.5\un{mm} and two
lenses with $R$= 0.2\un{mm}, all made out of high-purity Beryllium
(Be). The use of the condenser-CRLs improved the optical efficiency 
of the system (absorption of X-rays in the condenser-CRLs was only about 6\,\%.), as it increased the flux on the imaged sample area.   The divergence of the photon beam is not altered significantly, as the condenser CRL works almost in a 1:1 magnification geometry. A flux of approximately $2 \cdot 10^{12} \un{photons/s}$ was
incident to the sample. The sample was mounted on a six circle
diffractometer. The scattering plane coincided with the horizontal
plane.

The detector consisted of a scintillator screen, magnifying optics,
and a high resolution CCD-camera. The 9.9\un{\mu m} thick LAG:Eu
scintillator on a 170\un{\mu m} YAG substrate converted X-rays into
visible light, which was projected onto the CCD by the objective lens
(Olympus UPLAPO $\times 10$, numerical aperture 0.4). The CCD camera
(pco SensicamQE) had $1376 \times 1040$ pixels (px) of size
$6.45\un{\mu m/px} \times 6.45\un{\mu m/px}$ and 12 bit depth,
yielding a field of view on the scintillator of $887 \times
670\un{\mu m^2}$ with an effective resolution of 1.3\un{\mu m}. Each
CCD pixel imaged an area of $0.645\times 0.645\un{\mu m^2}$ on the
scintillator.

In front of the detector, on the same diffractometer arm, a second set
of paraboloid Be CRLs (66 lenses with apex-radius of curvature $R$=$50\un{\mu m}$) was
mounted as X-ray objective lens, i.e.~to image the diffracted
intensity pattern at the sample exit surface onto the detector. These
lenses were mounted on translation and rotation stages to align the
lens stack, in particular to tune the sample-to-lens distance to
achieve best focusing onto the detector.

The focal length of this lens stack at 11\un{keV} was about
14\un{cm}, so that a $\approx 4$-fold magnified image was achieved
with the lens center placed about 18\un{cm} downstream of the
sample. The effective aperture was about 240\un{\mu m}, giving a
corresponding diffraction limit of 130\un{nm}. The transmission through the lens stack is reduced by the absorption from the thinnest lens part, plus the increased absorption for rays travelling further away from the lens center, resulting in an effective aperture with Gaussian profile \cite{Snigirev97,lengeler99}. The first contribution is easy to calculate and gives an absorption of 18\,\%. Considering the size of the illuminated sample ($\approx$ 200\,nm, see below) and approximating the reflected beam as a parallel beam, the total absorption is closer to 50\,\%. 

Scaling the effective detector resolution by the magnification factor
4 to $1.3\un{\mu m}/4 = 0.33\un{\mu m}$, we expected a resolution
limit of $\sqrt{(0.33\un{\mu m})^2+(130\un{nm})^2} \approx 350\un{nm}$
with this set-up.

\setlength\fboxsep{0pt}
\setlength\fboxrule{0.2pt}
\begin{figure}[htbp!]
  \begin{center}
    \newcommand{\mywidth}{0.25\columnwidth}               % ###FOR ONE COLUMN
    \begin{tabular}{ll}
      (a) & (b)\\
      \includegraphics[width=\mywidth,clip=true,angle=180]{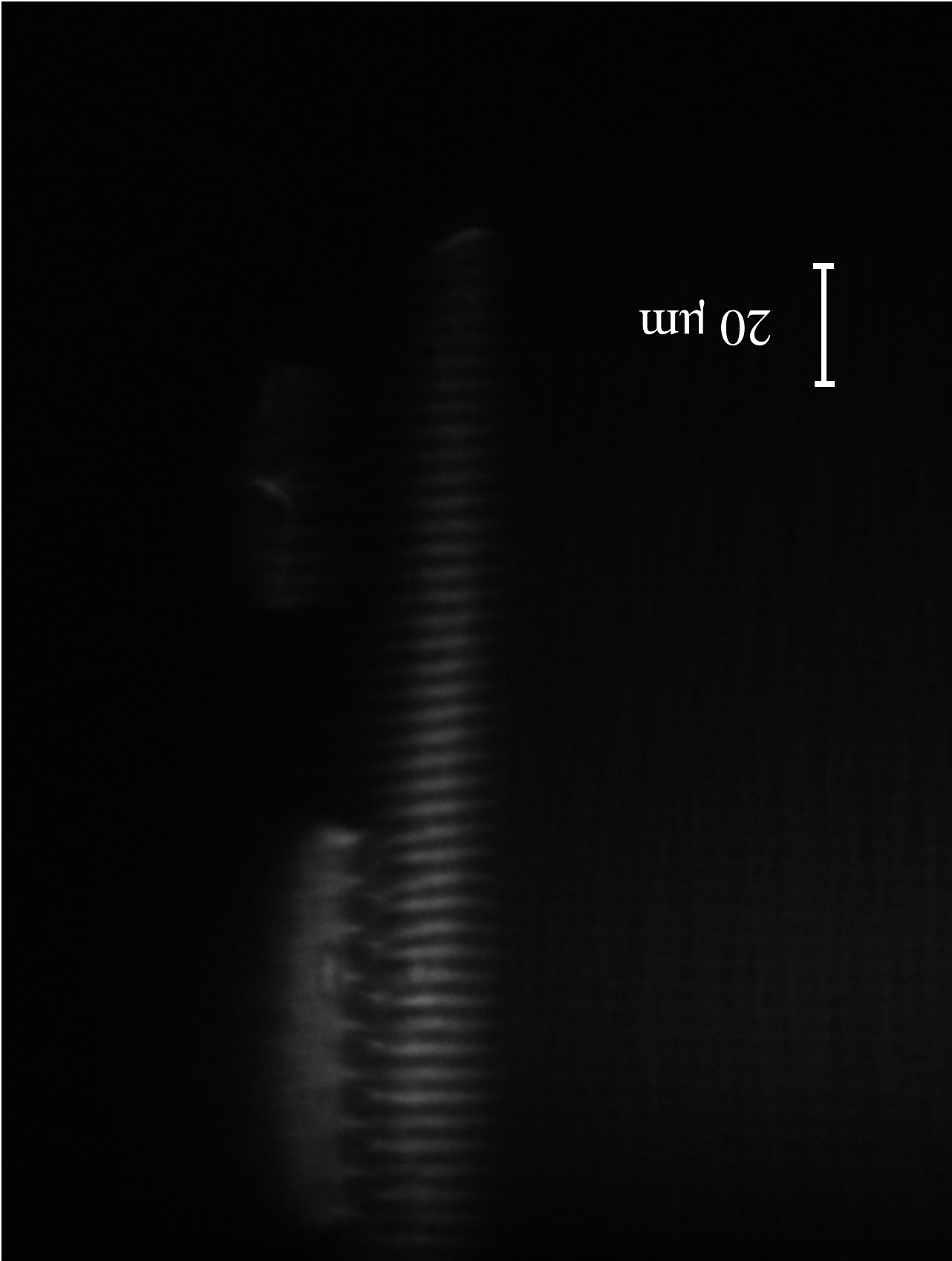}
      &
      \includegraphics[angle=-90,width=\mywidth]{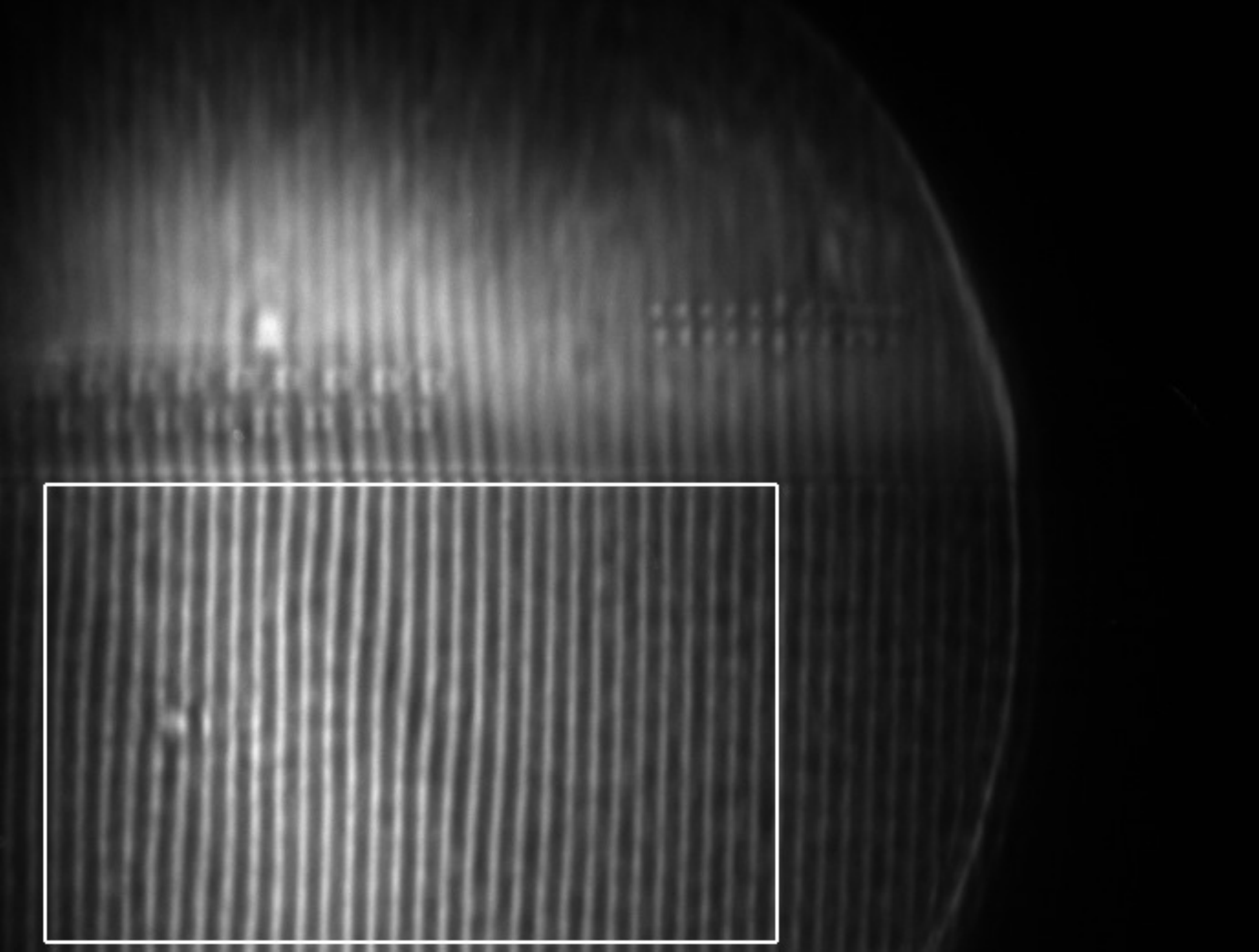}\\
      (c) & (d)\\
      \includegraphics[angle=-90,width=\mywidth]{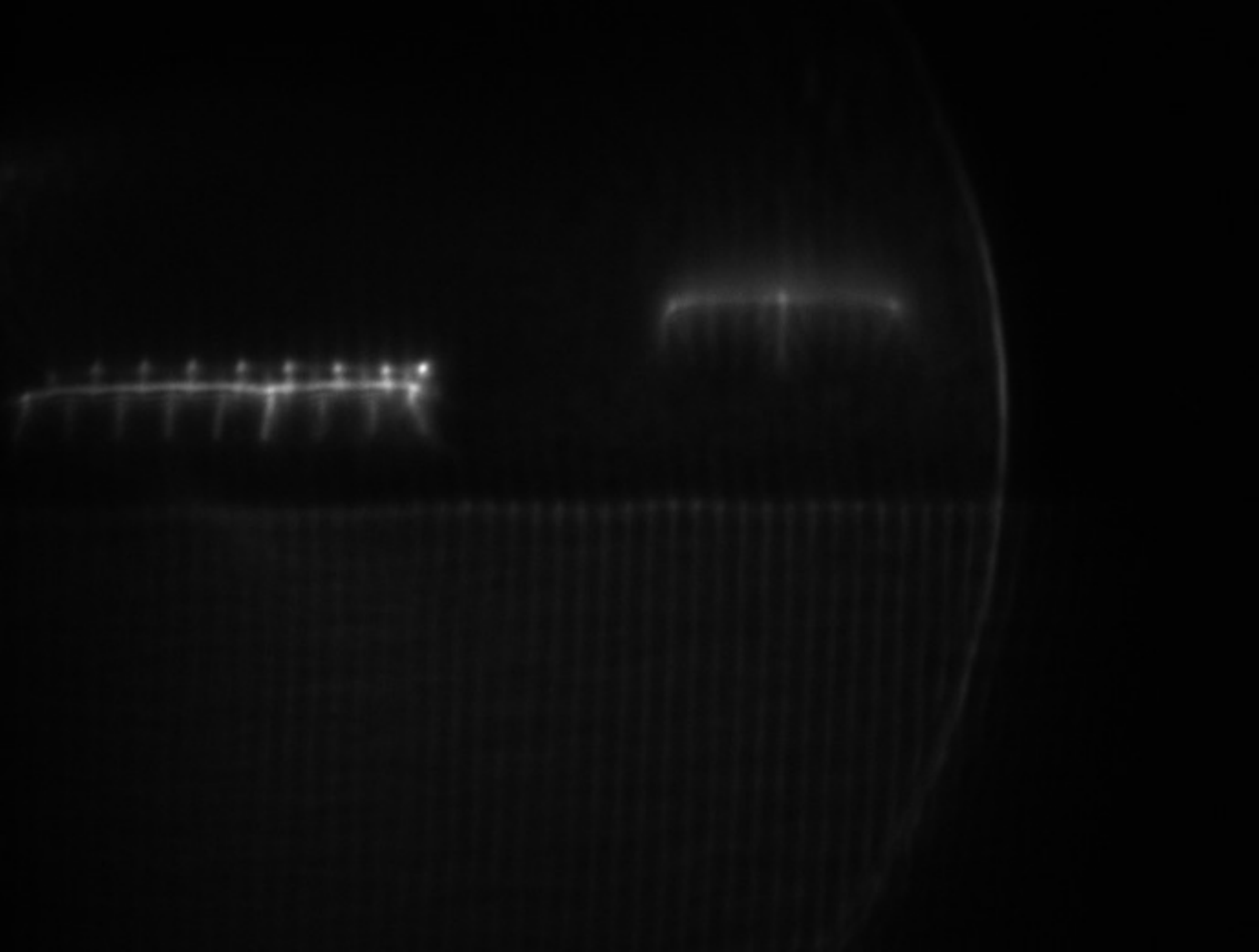}
      &

      \includegraphics[angle=-90,width=\mywidth]{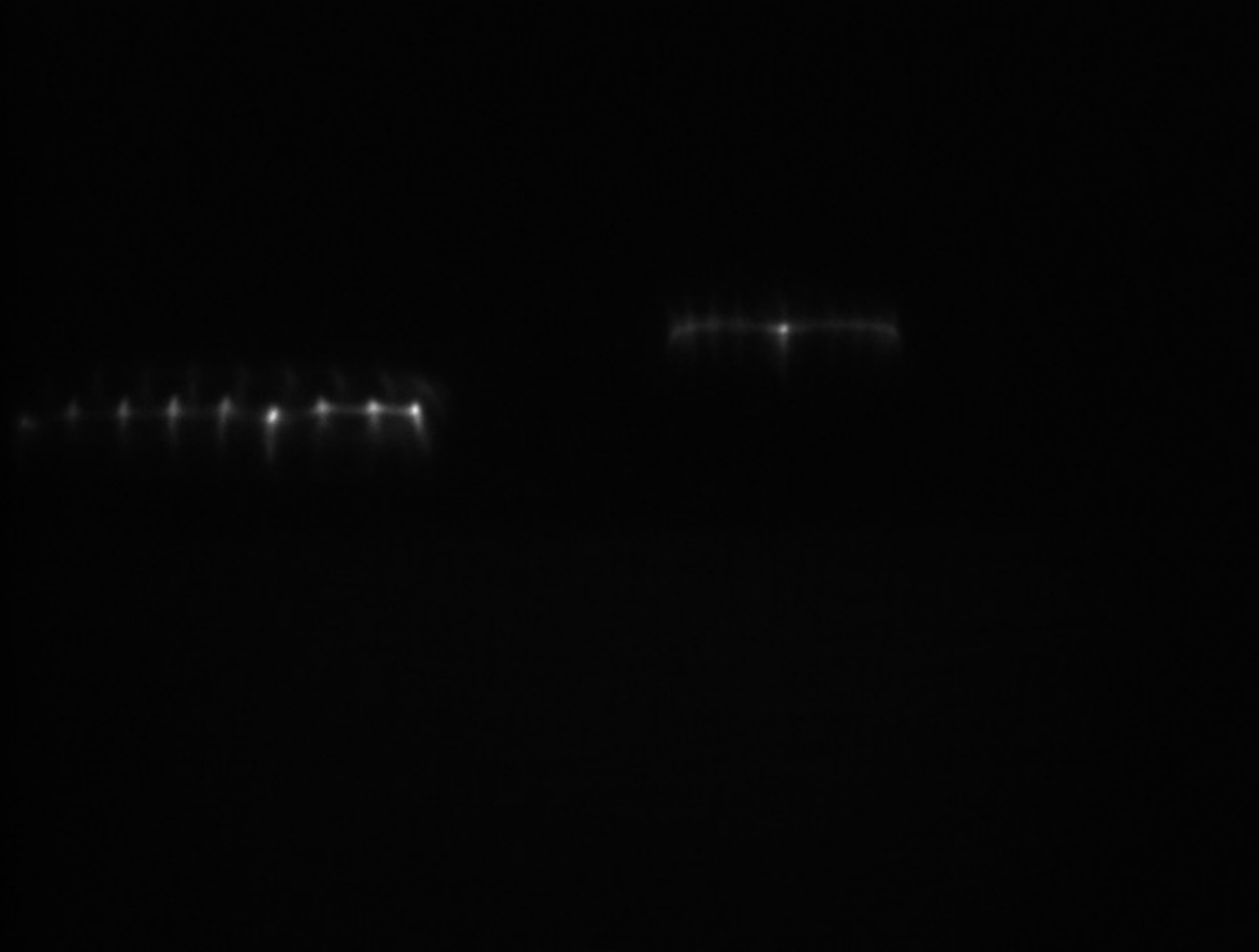}\\
      (e) & (f) \\
      \imagetop{\includegraphics[angle=0,width=\mywidth]{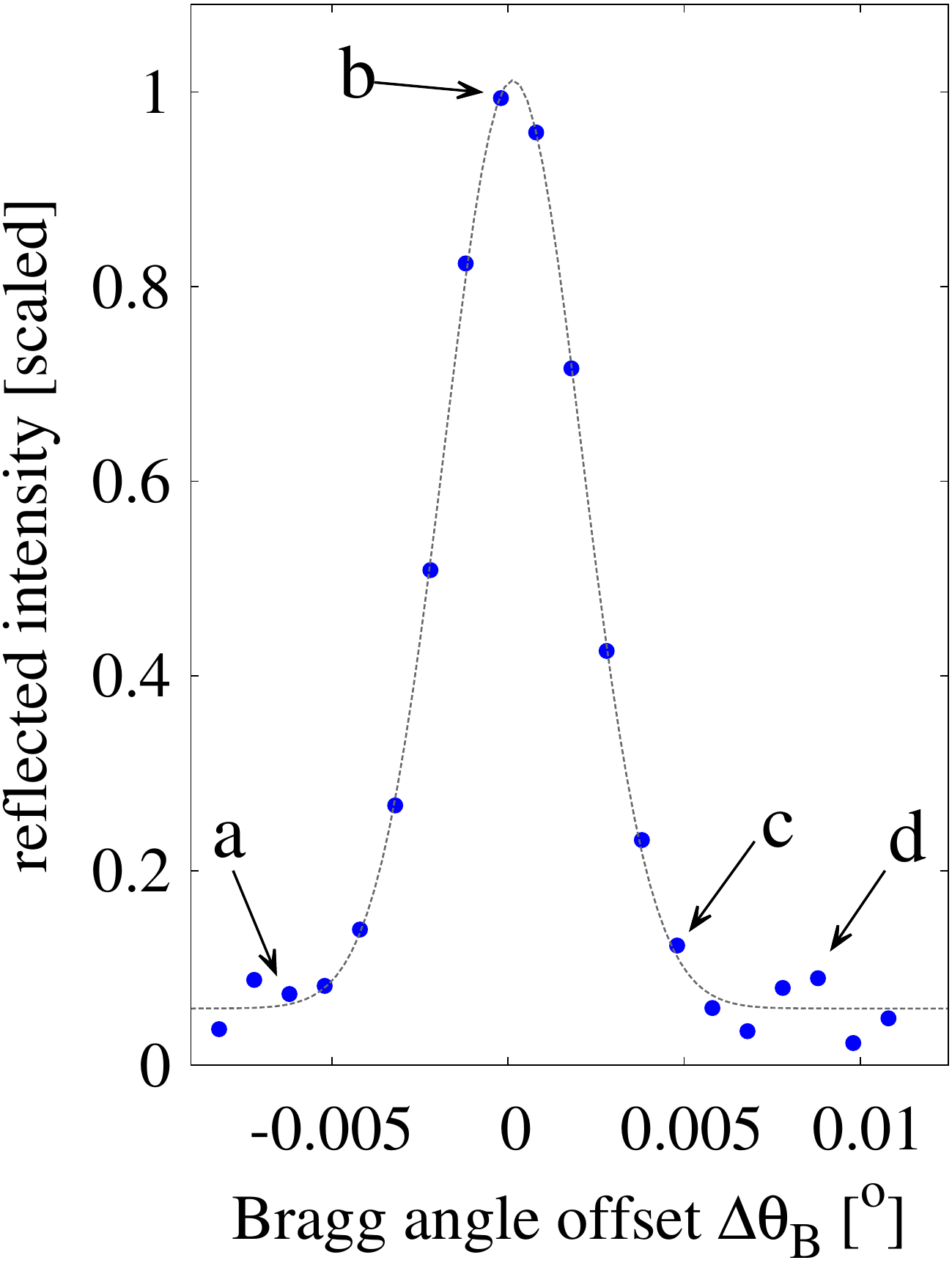}}
      &
      \imagetop{\includegraphics[angle=90,width=\mywidth]{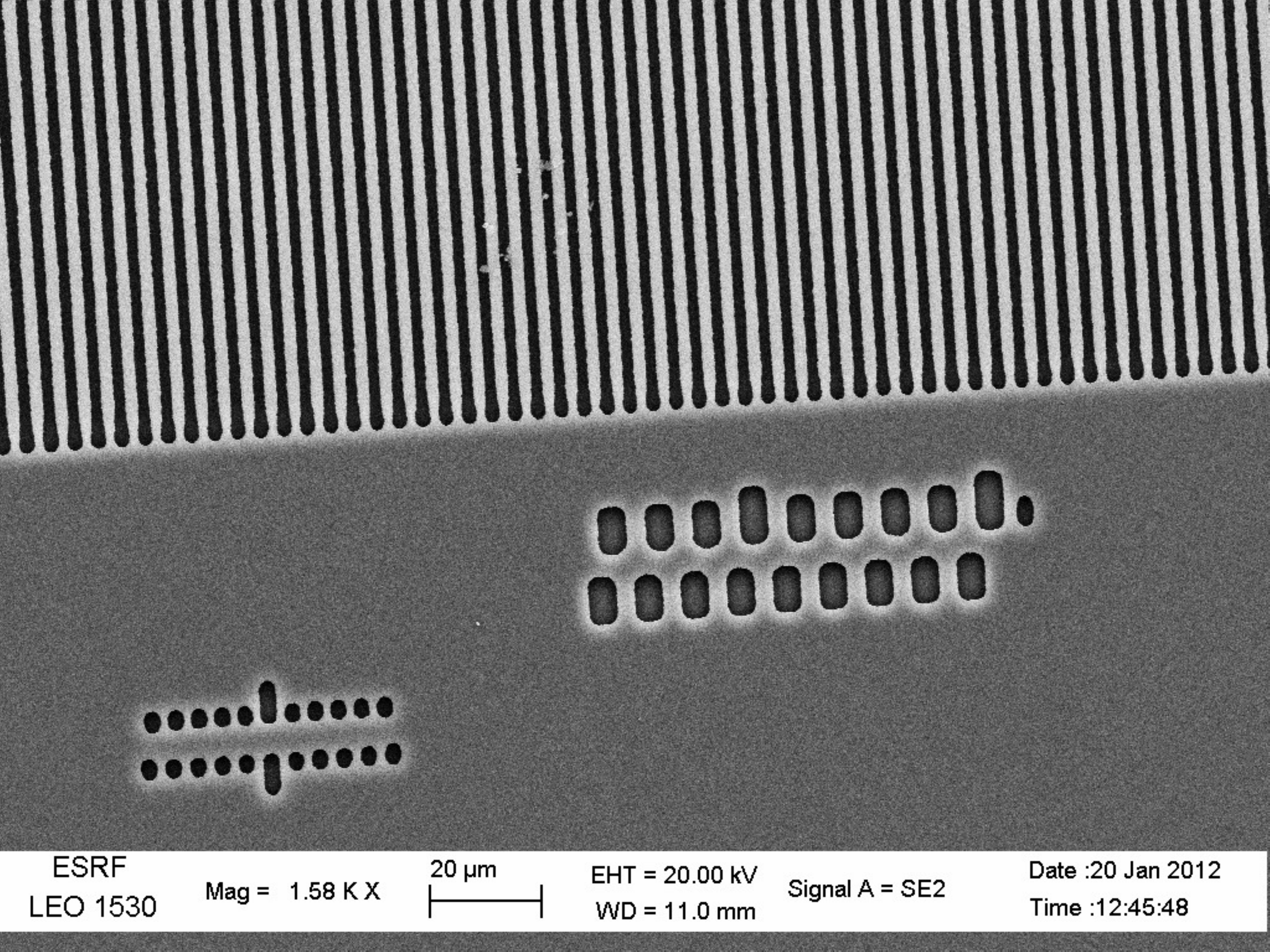}}\\
    \end{tabular}
  \end{center}
  \caption{\label{fig:stripe_results} Diffraction microscopy images
    (exposure time 2.5\un{s}) of the \chem{SiO_{2}} stripe structure
    at different Bragg angles: a) at $0.006^\circ$ below the maximum
    diffracted intensity; b) almost at the maximum; c) $0.005^\circ$
    above the maximum; d) $0.009^\circ$ above the maximum.  The beam 
    travels from left to right. e) Rocking
    curve as measured by a photo diode, indicating also the angle 
    positions corresponding to Figs. \ref{fig:stripe_results}a) to \ref{fig:stripe_results}d). f) Scanning electron
    microscope image of the same \chem{SiO_{2}} stripe system. Note that b) shows stripe like intensity in a region where e) shows a homogeneous \chem{SiO_{2}} surface. This indicates a strain propagation in the Si beyond the etched areas.}
\end{figure}

Two samples were imaged in Bragg geometry. The first sample was a Si
(111) wafer upon which a regular stripe pattern of amorphous
$\mathrm{SiO_2}$ has been fabricated by thermal oxidation followed by
standard photo-resist etching. The $\mathrm{SiO_2}$ layer was
$z=1.15\un{\mu m}$ thick, and was etched to fabricate 2\un{\mu m} wide
windows with a period of 4\un{\mu m}. The substrate was aligned in the
diffractometer to set the oxide stripes parallel to the diffraction
plane. In order to record magnified diffraction images of the sample
the Bragg (333) reflection of the Si substrate (Bragg angle
$\theta_B=32.63^\circ$) was used.

The second sample was a linear Bragg-Fresnel lens (BFL) fabricated on
a Si (111) substrate (for the fabrication process see
\cite{aristov87}). The basic geometrical parameters were: an
outermost zone width of 0.5\un{\mu m}, a height of the structure of
4.4\un{\mu m}, and an aperture of 200\un{\mu m}. Again, the sample was
aligned with the structures parallel to the horizontal scattering
plane. Again, the Si (333) ($\theta_B= 32.63^\circ$) reflection was
studied.

\section{Resolution}

The homogeneous periodicity of the $\mathrm{SiO_2}$ line pattern
(Fig.~\ref{fig:stripe_results}) was used to calibrate the effective
magnification of our configuration. The mask used to produce the
pattern had a period of 4\un{\mu m}, in good agreement with the value,
4.1(1)\un{\mu m}, obtained by scanning electron microscopy (SEM), see
Fig.~\ref{fig:stripe_results}f. Our X-ray image of this structure
shows 15 periods over 395(5)\un{px}. (Fig.~\ref{fig:stripe_results}b)
The line spacing on the fluorescence screen of the detector was
therefore $0.645\un{\mu m/px}\cdot 395\un{px}/15 = 17.0(2)\un{\mu m}$,
yielding a magnification factor of $17.0\un{\mu m}/4.1\un{\mu
m}=4.2(1)$ for the CRL stack and $4.1\un{\mu m}\cdot 15/395\un{px} =
0.156(4)\un{\mu m/px}$ for the overall experiment. The resulting field
of view on the sample was $\approx 162\un{\mu m}/\sin(\theta_B)$ in
the horizontal (within the scattering plane) and $215\un{\mu m}$ in
the vertical direction (perpendicular to the scattering plane).

\begin{figure}[ht!]
  \centerline{
    \includegraphics[width=0.6\columnwidth,trim=20mm 2mm 30mm 15.5mm, clip=true]{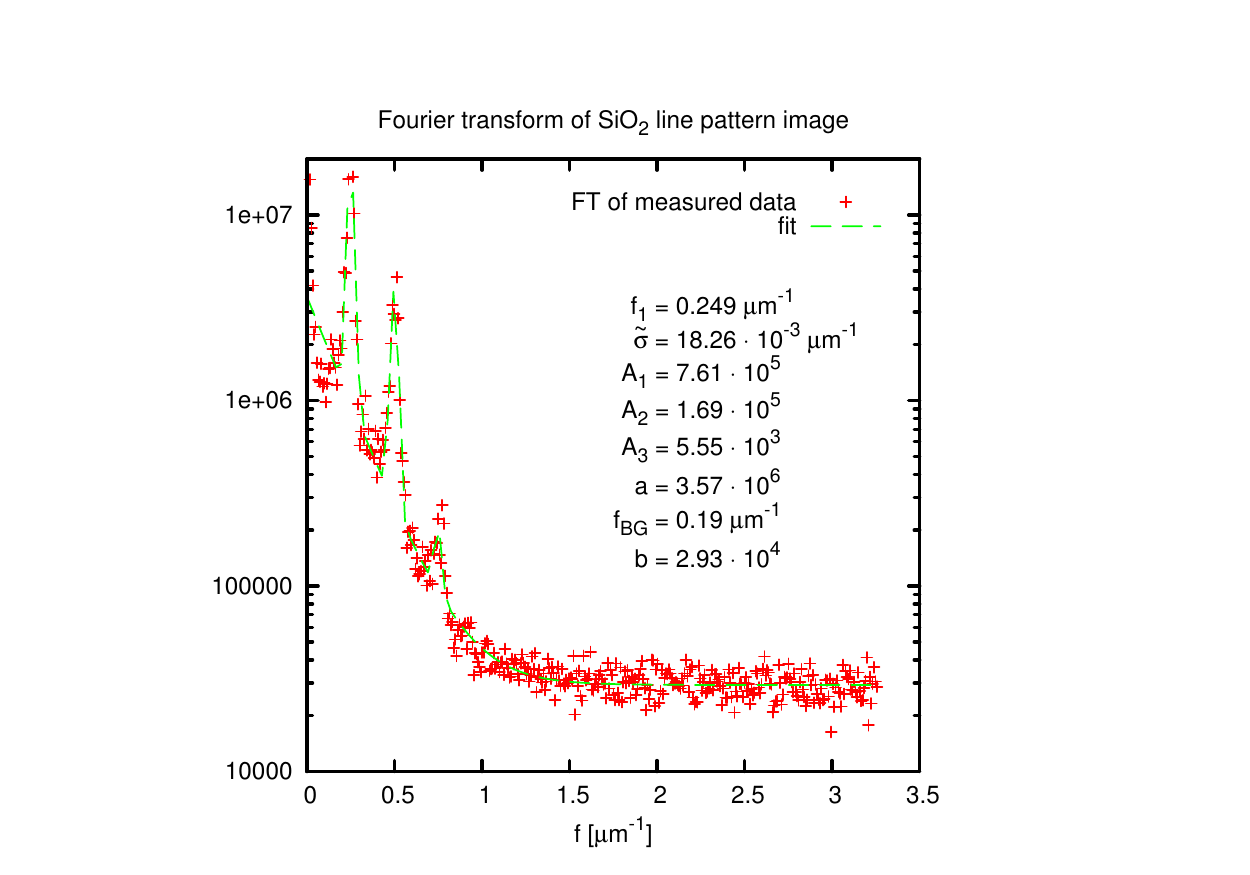}
  }
  \caption{\label{fig_ft_sio2}Fourier transform of the diffraction
    microscopy image shown in Fig.~\ref{fig:stripe_results}(b). The
    region of interest was divided into 10 vertical slices. The data
    shown here represent the magnitude of the Fourier transform
    averaged over these 10 slices. The parameters listed are the
    result of a fit to eq.~\ref{eq_model_sio2}.}
\end{figure}

An upper limit for the effective resolution of our imaging system can
be estimated from the Fourier transform (FT) of the image (see
Fig.~\ref{fig_ft_sio2}). Peaks corresponding to the fundamental,
second and third harmonics of the structure are clearly visible above
a two-component background. For quantitative analysis, the FT was
fitted to a model function
\begin{equation}
  \tilde{I}(f) = \sum\limits_{n=1}^{3} \left( A_n \cdot g(f-n f_1) \right)
  + a e^{-f/f_\mathrm{BG}} + b,
  \label{eq_model_sio2}
\end{equation}
where $\tilde{I}(f)$ is the Fourier transform of the image at spatial
frequency $f$. The background is composed of a constant and an
exponentially decaying term with characteristic frequency
$f_{\mathrm{BG}}$. The harmonics are modelled by Gaussians $g(f) =
(2\pi \tilde{\sigma}^2)^{-1/2} \exp(-f^2/2\tilde{\sigma}^2)$ with
amplitude factors $A_n$ for the $n$-the harmonic. The magnification
factor determined above (0.156\un{\mu m/px}) was used to scale the
frequency axis. The resulting parameters are shown in
Fig.~\ref{fig_ft_sio2}. The ratio $A_1/A_3$ can be used to estimate
the modulation transfer function (MTF),
$\tilde{I}(f)=\tilde{c}(f)\cdot \mathrm{MTF}(f)$, where $\tilde{c}(f)$
is the FT of the scattering amplitude of the sample. For an ideal
square wave, $\tilde{c}(f_1)/\tilde{c}(3 f_1) = 3$. For a more smooth
modulation, e.g. resulting from continuous buckling of the Bragg
planes due to strain \cite{Kuznetsov04}, the higher harmonics will be
suppressed, $\tilde{c}(f_1)/\tilde{c}(3 f_1) > 3$ so that
\begin{equation}
  \frac{
    \tilde{I}(f_1)
  }{
    \tilde{I}(3 f_1)
  }
  =  
  \frac{
    \tilde{c}(f_1)
  }{
    \tilde{c}(3 f_1)
  }
  \cdot
  \frac{
    \mathrm{MTF}(f_1)
  }{
    \mathrm{MTF}(3 f_1)
  }
  \geq
  3 
  \cdot
  \frac{
    \mathrm{MTF}(f_1)
  }{
    \mathrm{MTF}(3 f_1)
  }
\end{equation}
The presence of a second harmonic at $2 f_1$ indicates that the
contrast does not follow an ideal square modulation.

In the absence of any further information, we assume the MTF to be a
Gaussian with standard deviation $\tilde{\sigma}_{\mathrm{MTF}}$, so
that $\tilde{\sigma}_{\mathrm{MTF}} = 2 f_1
\left(\log[\mathrm{MTF}(f_1)/\mathrm{MTF}(3f_1)]\right)^{-1/2}$. Using
the values obtained from the fit we find
$\tilde{\sigma}_{\mathrm{MTF}} \geq 0.254 \un{\mu m^{-1}}$,
corresponding to a Gaussian point spread function (PSF) with standard
deviation $\tilde{\sigma}_{\mathrm{PSF}} = {1}/(2\pi
\tilde{\sigma}_{\mathrm{MTF}}) \leq 0.625\un{\mu m}$ and full width at
half maximum (FWHM) $\leq 1.47\un{\mu m}$ on the sample (9.4\un{px} on
the CCD).

\begin{figure}[ht!]
  \begin{center}
    \newcommand{\mywidth}{0.35\columnwidth}
    \begin{tabular}{ll}
      (a) & (b)\\
      \includegraphics[width=\mywidth, clip=true]{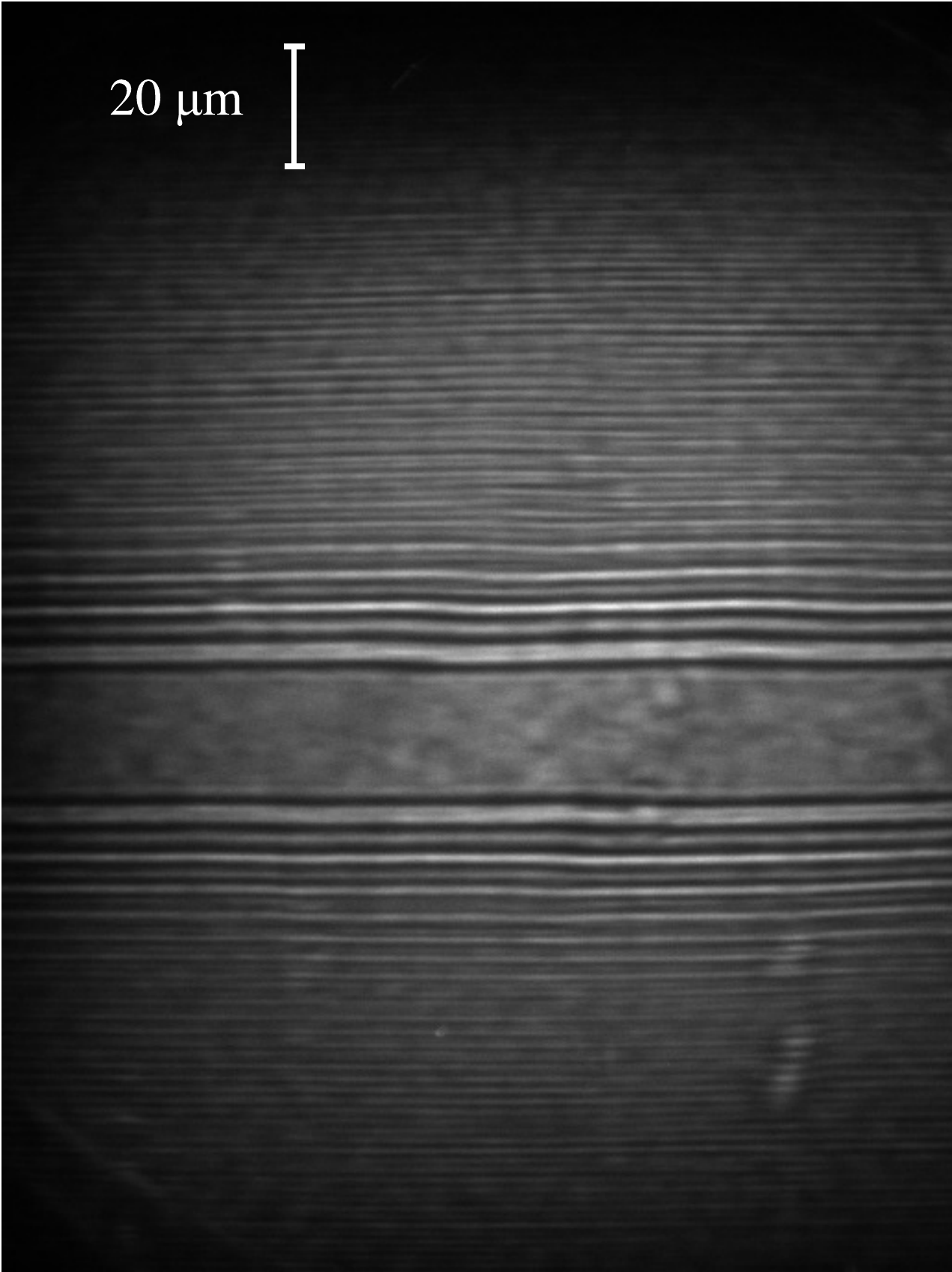}
      &
      \includegraphics[angle=90,width=\mywidth]{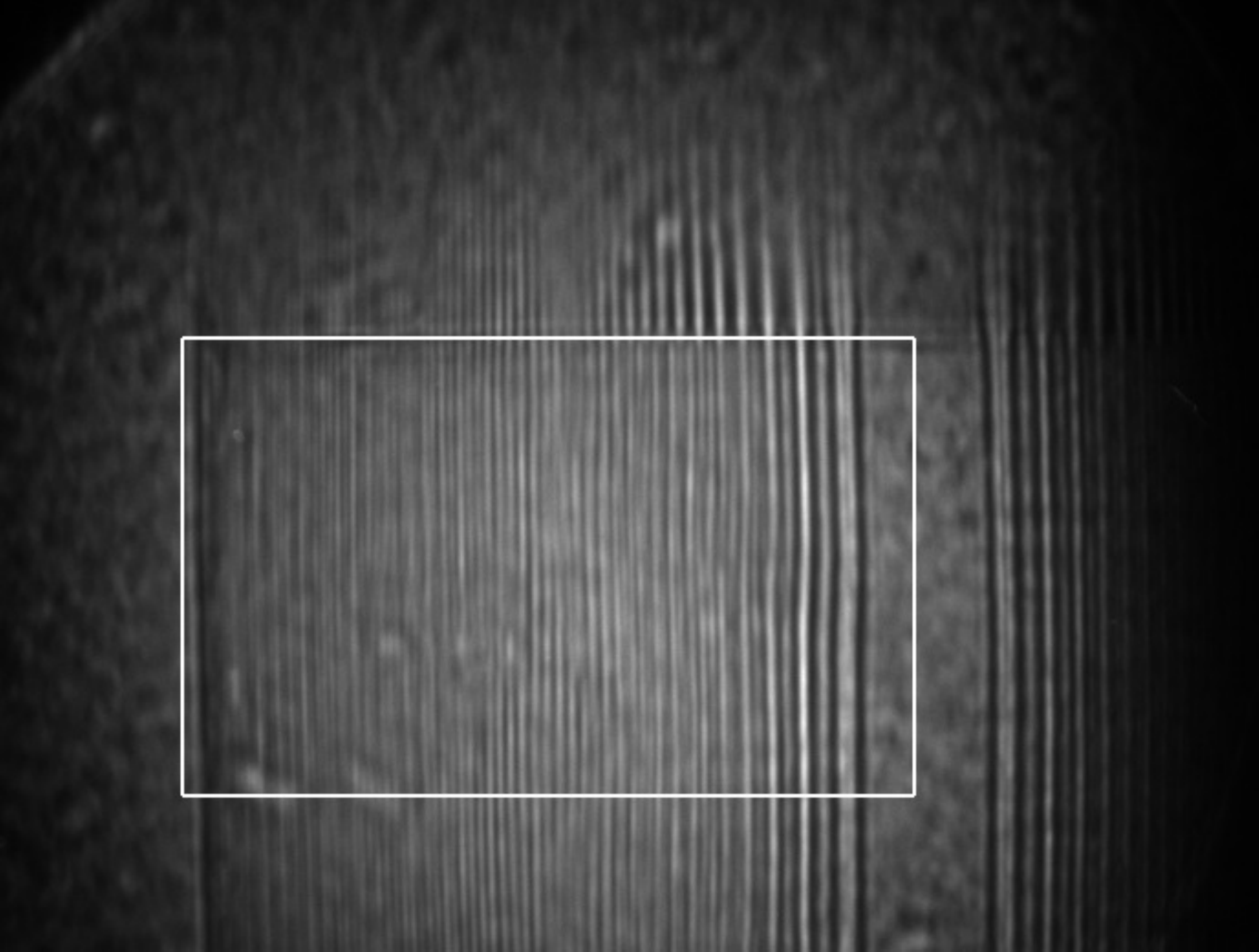}\\
      (c) & (d)\\
      \imagetop{\includegraphics[width=\mywidth]{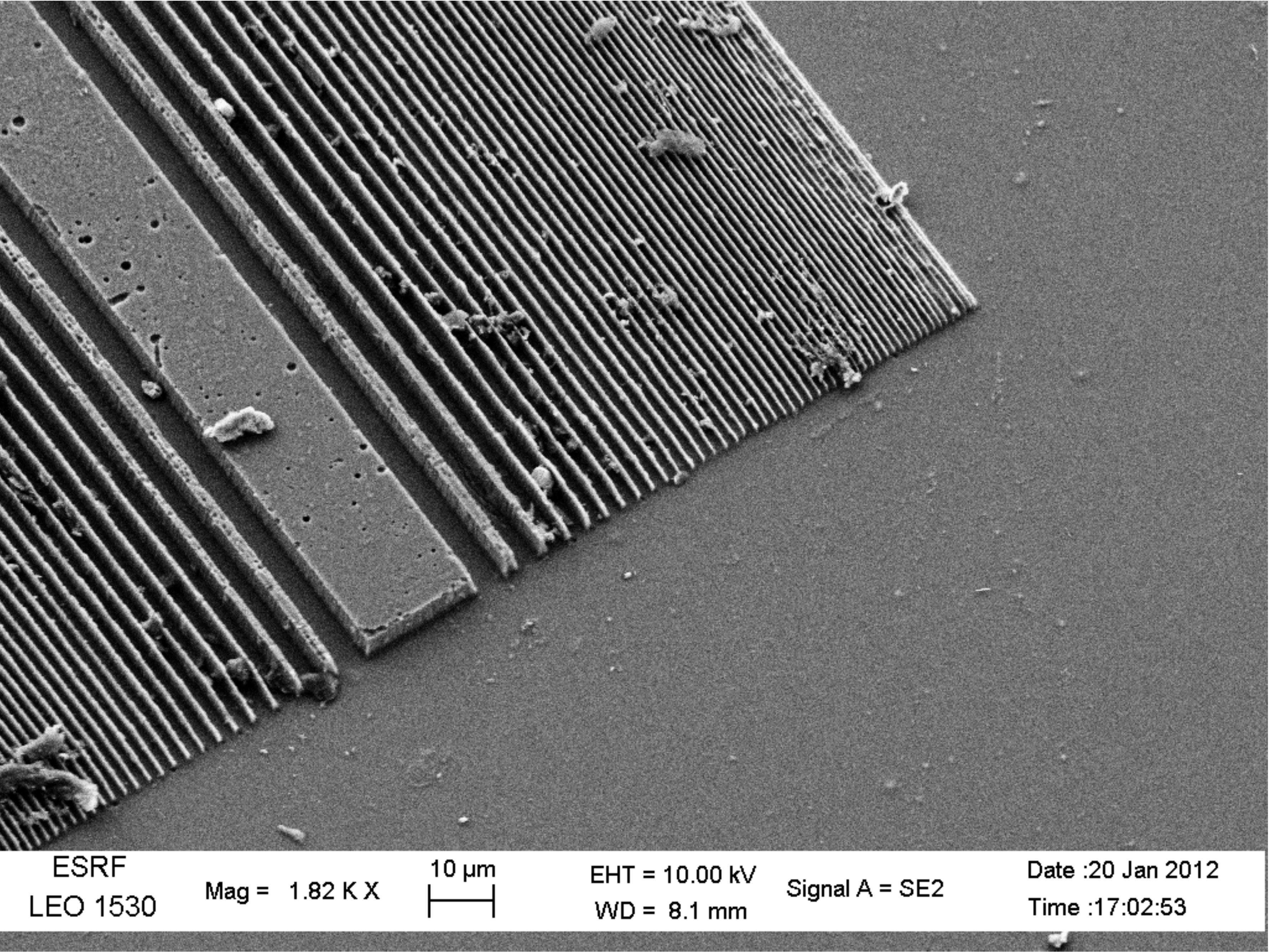}}
      &
      \imagetop{\includegraphics[width=\mywidth,clip=true,trim=50 5 90 90]{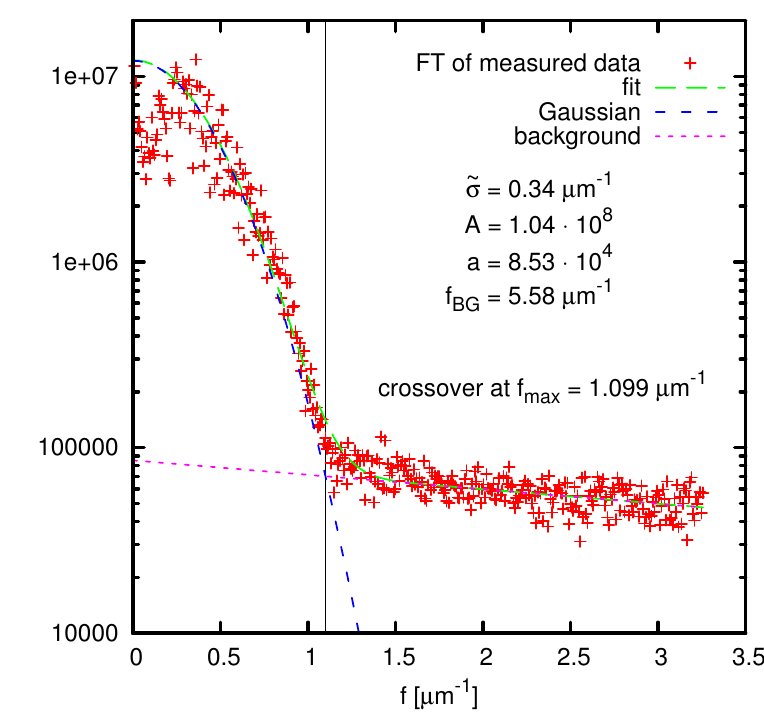}}
    \end{tabular}
  \end{center}
  \caption{\label{fig_zone_plate} a), b) Diffraction microscopy images of a
    Bragg-Fresnel lens (Exposure times a) 1\un{s} and b) 2.5\un{s}).
    c) Scanning electron microscope image of the same Bragg-Fresnel
    lens. d) Fourier analysis of the stripe structures. The region of
    interest (ROI) shown in (d) was divided into 10 vertical
    slices. Each slice was Fourier transformed. The average of the
    resulting magnitudes were fit to eq.~\ref{eq_model_bfl}. The
    modulation transfer function intersects the background at
    $f_{\mathrm{max}}=1.10\un{\mu m^{-1}}$, indicating a minimum
    observable peak-to-valley distance of
    $1/(2f_{\mathrm{max}})=0.46\un{\mu m}$.  }
\end{figure}

Further experiments were performed on a Bragg-Fresnel lens. In
this sample, the zone width decreases away from the center, thus
yielding a richer Fourier spectrum that should provide more detailed
information on the MTF and the effective resolution of our system.

Fig.~\ref{fig_zone_plate} shows the Bragg microscopy images of the Bragg-Fresnel lens
(panels a and b), a Scanning electron microscopy image of the same
structure (panel c), and a Fourier analysis (panel d) of the region of
interest shown in panel b. 

As above, Fourier analysis was performed to reveal the resolving power
of our experiment. The region of interest shown in panel b) was
divided into 10 vertical slices. Each slice was Fourier transformed,
The average magnitude of the FT is shown in panel d). A Gaussian MTF
with an exponential background was fitted to the average FT, again
using the magnification factor of 0.156\un{\mu m/px} to scale the
frequency axis.
\begin{equation}
  \tilde{I}(f) = A \cdot g(f) + a \cdot e^{-f/f_{\mathrm{BG}}}
  \label{eq_model_bfl}
\end{equation}
From the resulting parameters (listed in Fig.~\ref{fig_zone_plate}d) two estimates of the
resolution were derived: The standard deviation of the MTF, and the
frequency $f_{\mathrm{max}}$ where the MTF falls below the background.

The fit yielded $\tilde{\sigma}_{\mathrm{MTF}}=0.34\un{\mu m^{-1}}$
and $f_{\mathrm{max}}=1.10\un{\mu m^{-1}}$. The period of a structure
at this frequency is
$\lambda_{\mathrm{min}}=1/f_{\mathrm{max}}=0.92\un{\mu m}$, so that
the smallest observable peak-to-valley valley distance on the sample
is $\lambda_{\mathrm{min}}/2 = 0.46\un{\mu m}$. The standard deviation
of the MTF corresponds to a PSF with standard deviation
$\sigma_{\mathrm{PSF}}=1/(2\pi
\tilde{\sigma}_{\mathrm{MTF}})=0.47\un{\mu m}$ and FWHM 1.1\un{\mu m}
on the sample (7.1\un{px} on the CCD), slightly better than the
estimate obtained above for the \chem{SiO_2} stripe pattern.

We have thus shown that our setup reaches sub-micrometer resolution in
the vertical direction (perpendicular to the scattering plane). In the
horizontal direction (parallel to the scattering plane) the sample
does not lie perpendicular to the camera plane. Viewing the sample at
the Bragg angle $\theta_B$ yields a projected in-plane images size
that is smaller by a factor $\sin(\theta_B)$, here $\sin(32.63^\circ)=
0.539$. Assuming that the resolution limit is given by the detector
and the diffraction limit of the imaging lenses, the in-plane
resolution at the sample surface is then degraded by the factor
$1/\sin(\theta_B) = 1.85$ compared to the vertical out-of-plane
direction. Furthermore, within the scattering plane, the beam
transverses the scatterer partly. The beam path length inside the
sample (limited by absorption or extinction) is comparable to the
resolution, so that additional blurring is to be expected.  This could
be avoided in backscattering geometry ($\theta_B \approx 90^\circ$),
or when imaging high-Z materials with low penetration depth.

\section{Discussion}

Using the example of the \chem{SiO_2} stripe sample, we now discuss the 
information obtainable from diffraction images taken at different Bragg 
angles. We recall that the reflected intensity stems form the underlying
single crystalline Si waver and not from the amorphous top structure of 
 \chem{SiO_2} stripes.

Contrast in the diffraction image may arise from several effects: (a)
absorption leading to different amplitudes of rays that do or do not
travel through the thin \chem{SiO_2} layer (b) phase shift between
these beams, as the index of refraction of \chem{SiO_2} is different
from unity, and (c) local variations of the Si(111) reflectivity due
to strain in the Si substrate induced by the overlying \chem{SiO_2}
layer.

For (a) and (b) we can calculate the expected contrast. The index of 
refraction of \chem{SiO_2} at $E=11\un{keV}$
is $n=1-\delta+i \beta$ with $\delta=3.8\cdot 10^{-6}$ and $\beta=
2.7\cdot 10^{-8}$ \cite{CXROweb}, assuming a density of
2.2\un{g/cm^3}. The path length of the X-rays through
\chem{SiO_2} is $L=2z/\sin(\theta_B)=4.3\un{\mu m}$. The E-field
amplitude of the beam travelling through the \chem{SiO_2} layer is therefore reduced to
$\exp(-2\pi L \beta/\lambda) = 0.9935$, whereas its phase is
shifted by $L\delta/\lambda\cdot 360^\circ=52^\circ$ as compared to the beam travelling through an adjacent groove via the bare Si surface. Consequently,
the absorption contrast (a) is expected to be 1-$\frac{0.9935^2}{1}$ = 1.3\,\%. The phase contrast (b) occurs through
interference at the edges. It can be estimated in calculating the intensity resulting from the superposition of two beam parts that travel (i) through the SiO$_2$ and acquiring the 52$^\circ$ phase shift and a part (ii) that does not travel through the SiO$_2$, but through the groove. This intensity is to be compared with the signal from two beams that did not experience a relative phase shift. We obtain a phase contrast of 1-$\frac{|1+\exp(i\cdot 52^\circ)|^2}{|1+1|^2}$ = 20\,\%. For the regular \chem{SiO_2} pattern of 
Fig \ref{fig:stripe_results}b), the measured contrast was  35\,\%. So a part of the contrast must come from strain. 

The magnitude of the strain contrast (c)
is difficult to estimate. However, for (a) and (b) the contrast at
each point of the rocking curve should be identical, whereas strain
might shift and broaden the rocking curve, so that for (c) the
contrast in the diffraction micrographs at different points of the
rocking curve might differ \cite{Tanuma06}.

This can indeed be seen when comparing images recorded at different
positions on the rocking curve (Fig.~\ref{fig:stripe_results}a--d). Fig.~\ref{fig:stripe_results}a and \ref{fig:stripe_results}b show intensity
on the right-hand side of the \chem{SiO_2} line pattern, caused by
strain propagation beyond the etched areas. Fig.~\ref{fig:stripe_results}c and~d show that
control structures etched into the \chem{SiO_2} (shown in the right
part of the SEM image, Fig.~\ref{fig:stripe_results}f) cause strong strain in the \chem{Si}
substrate.

Such shifts of the rocking curve can occur via two routes, local
tilting of the lattice planes, or local modifications of the lattice
parameter \cite{Aristov92,Aristov92b}. In the former, a positive tilt
on one side of a straining feature should be accompanied by a negative
tilt on the opposite side. The corresponding areas should be visible
at angles symmetric to the center of the rocking curve. A local
modification of the lattice parameter, on the other hand, would lead
to a unidirectional shift with respect to the unstrained rocking curve
\cite{Tanuma06}.

The sharp features visible in Fig.~\ref{fig:stripe_results}d appear only $\approx
0.009^\circ$ above the rocking curve, indicating that the lattice
parameter is compressed by 
\begin{equation}
	\frac{\Delta d}{d} 
	\approx \mathrm{cot}(\theta_B)  \cdot \Delta\theta
	\approx 2.5\cdot 10^{-4}.  
\end{equation} 

As shown in Fig.~\ref{fig:stripe_results}e, this strain level is clearly resolved in our
experiment. The sensitivity to lattice strain could be further
improved by selecting higher order Bragg reflections with narrower
rocking curves.

\section{Conclusion}

X-ray diffraction microscopy combines the advantages of X-ray
microscopy in forward scattering geometry and conventional diffraction
topography without image magnification:

\begin{itemize}
\item As in transmission X-ray microscopy, the effective resolution
  is greater than that achievable with conventional diffraction
  topography, which is limited by detector resolution and
  sample-to-detector distance (diffraction effects).
\item As in diffraction topography, the technique is sensitive to
  microscopic crystallographic imperfections such as strain,
  dislocations, twinning, etc.
\end{itemize}

As we have shown here, data acquisition is fast: A single exposure is
sufficient and contains all the maximum resolution information, thus
the technique is robust with respect to instabilities of the
experimental setup, and it has the potential to study transient,
non-equilibrium phenomena where it is impossible to acquire several
images of the same state.

It should be underlined that the proposed diffraction microscopy
technique has great potential for non-destructive studies of highly
deformed metals and alloys. By comparing the images taken at different
angular settings the contrast due to strain or orientation are easily
distinguished \cite{Afanasev71}. Adding a tomography option
($180^\circ$ rotation) will provide 3D mapping of the orientation and
strain of individual grains in polycrystalline materials.

The use of
CRLs is of particular interest in diffraction topography since these
optics are well adapted to focusing hard X-rays, and are relatively
straight-forward to implement on existing diffractometer setups.
The technique, however, can also be used with other imaging
systems such as Fresnel zone plates \cite{Tanuma06,Fenter06} or
mirrors such as Wolter optics \cite{Wolter52,Takano02}. This
flexibility enables the use of X-ray diffraction microscopy over a
very wide range of photon energies from sub-keV soft X-rays, e.g.~for
the study of multilayers, to very hard X-rays with several tens of
keV.  
Furthermore, the technique can be combined with other standard
X-ray techniques to access information unobtainable in transmission
geometry. Examples include grazing-incidence diffraction to image
micro- and nanostructures grown on a surface, magnetic scattering to
image ferromagnetic \cite{Kreyssig09} or antiferromagnetic
\cite{Lang04} magnetic domain patterns and the imaging of
ferroelectric domains \cite{Fogarty96}.

Finally, the field of view and the magnification can be adjusted
in-situ simply by changing the number of lenses (e.g.~by using a
Transfocator) and the sample-to-lens or lens-to-detector distance.

{\textbf{\\ Acknowledgments}}

We acknowledge
the European Synchrotron Radiation Facility (ESRF) for the provision
of beam time on ID06. C.~D.~thanks R.~Barrett for stimulating
discussions and critical reading of the manuscript.

\printbibliography

\end{document}